\documentclass[nonacm]{acmart}

\setcopyright{None}

\usepackage{graphicx}  
\usepackage{subcaption} 
\usepackage{csquotes}
\usepackage{siunitx}
\usepackage{paralist}
\usepackage{enumitem}
\begin{document}

\title{The Privacy Guardian Agent: Towards Trustworthy AI Privacy Agents}

%
%
\author{Vincent Freiberger}
\affiliation{%
  \institution{Center for Scalable Data Analytics and Artificial Intelligence (ScaDS.AI) Dresden/Leipzig, Leipzig University}
  \city{Leipzig}
  \country{Germany}
}

\email{freiberger@cs.uni-leipzig.de}

\begin{abstract}
The current "notice and consent" paradigm is broken: consent dialogues are often manipulative, and users cannot realistically read or understand every privacy policy.
While recent LLM-based tools empower users seeking active control, many with limited time or motivation prefer full automation. However, fully autonomous solutions risk hallucinations and opaque decisions, undermining trust.
I propose a middle ground — a Privacy Guardian Agent that automates routine consent choices using user profiles and contextual awareness while recognizing uncertainty. It escalates unclear or high‑risk cases to the user, maintaining a human‑in‑the‑loop only when necessary. To ensure agency and transparency, the agent's reasoning on its autonomous decisions is reviewable, allowing for user recourse. For problematic cases, even with minimal consent, it alerts the user and suggests switching to an alternative site. This approach aims to reduce consent fatigue while preserving trust and meaningful user autonomy.

\end{abstract}






\maketitle

\section{Motivation}
The current approach of "notice and consent" on the web is dysfunctional by design. It places an impossible cognitive burden on users who cannot read, let alone comprehend, every privacy policy they encounter~\cite{shiri,mcdonald2008cost}. Further, consent dialogues tend to employ dark patterns that manipulate users into habituated acceptance rather than informed choice~\cite{nouwens}. While recent Large Language Model (LLM) tools promise to restore informational sovereignty by summarizing and explaining policies~\cite{freiberger2026,sun2025empowering,chen2025clear}, they primarily serve motivated users seeking active control. For the majority, particularly "Unconcerned" or "Lazy Experts" privacy profiles~\cite{dupree2016privacy}, the friction of manual review remains a barrier, often exacerbated by time constraints and familiarity bias~\cite{shiri,freiberger2026}. However, fully automated agentic solutions risk eroding trust through hallucinations or misaligned, opaque decision-making~\cite{freiberger2025explainable}.
As a result, we face a dilemma: manual control is impractical for most, but full automation removes the user agency necessary for meaningful consent.

\section{A Privacy Guardian Agent}
I argue for a middle ground -- a hybrid system that combines user privacy profiles~\cite{dupree2016privacy,marky} with contextual analysis of data flows to automate routine consent decisions while escalating uncertain or high-risk cases to the user. 
Going beyond choosing consent options, the agent should help users decide whether to use a service at all when even "essential-only" consent implies unacceptable risks to users, like sensitive inferences or suspicious third-party sharing. The agent alerts the user and, inspired by PrivacyCheck~\cite{nokhbeh2020privacycheck}, proposes alternative services that better align with their privacy profile.

This approach minimizes consent fatigue while preserving user agency. Users can always review the agent's rationale for any decision and override it when needed.
By delegating routine decisions to the agent, users avoid habituation and retain cognitive resources and control for uncertain or high-stakes decisions, aligning with Zhang et al.'s (2026)~\cite{zhang2026scoping} design directions for context-aware just-in-time interventions.


\textbf{User Profiling: }New users disengage when faced with an extensive number of setup questions. Instead, utilizing privacy personas~\cite{marky,dupree2016privacy} along the dimensions of motivation and privacy knowledge can help calibrate the agent to users' needs. For example, an "Unconcerned" user might want to auto-accept analytics cookies on news sites, while a "Fundamentalist" profile would auto-reject third-party tracking. The tool could utilize a user self-assessment on motivation and privacy knowledge following Marky et al. (2024)~\cite{marky} for initial privacy profiling and potentially refine and adjust based on locally stored continued usage data. 

\textbf{Contextual Integrity analysis: }The agent should understand the context of a data request (like sensitivity of data, estimated urgency of users tasks or evaluation of stated data collection purposes) and leverage the obtained user privacy profile. Guardian Agents could utilize LLMs to parse privacy policies into Contextual Integrity norms~\cite{barth2006privacy}. So purposes for collected data types for a given service could be evaluated considering how privacy-sensitive users actually are. In case CI norms cannot be reliably extracted, the decision should be escalated to the user.

\textbf{Reliability Calibration: }To reduce the risk of AI hallucination or factual errors, I suggest reliability calibration. The system should provide privacy policy evidence to support its interpretation, making its reasoning transparent and auditable. Further, the agent should communicate the reliability of its risk assessment utilizing measures like model uncertainty, consistency checks or following more elaborate approaches like Tanneru et al (2024)~\cite{tanneru2024quantifying}. 
It should flag inconsistencies, missing information, or ambiguous language that prevent reliable Contextual Integrity analysis.
When the agent cannot confidently determine whether a data flow violates the user's profile, it escalates to the user.
The escalation message explains what is uncertain and why it matters for the user's privacy preferences. In cases where the agent identifies unacceptable risk, it also communicates its confidence level in that assessment, allowing users to calibrate their trust appropriately. 


\section{Discussion}
A key question is how far consent automation can legitimately extend under GDPR, which requires consent to be "freely given" and "specific." I argue the agent should be framed as a decision-support tool rather than a legal proxy: the user configures the profile, reviews high-stakes decisions, and can audit past choices. Future work should examine under which conditions such delegation enhances rather than undermines autonomy in practice, and how reliability calibration influences trust. 
Moreover, the question of accountability should be explored: Who bears responsibility if the agent makes mistakes or misinterprets a policy? Beyond hallucinations, the agent could introduce new manipulation risks if it is captured by vendor incentives or adversarial policies requiring guardrails like a transparent rationale and robustness measures.

Ironically, a Privacy Guardian Agent that protects user privacy requires collecting and processing sensitive behavioral data, which creates a new privacy risk. User privacy profiles combined with browsing context and interaction patterns could reveal details about users' values, concerns, and online habits. Consequently, profile data must be stored locally and processed either locally or with privacy-preserving techniques like privacy sanitization in place. The agent should retain only aggregated preference patterns, not granular histories. Without such safeguards, Guardian Agents risk becoming data protection risks themselves, undermining the informational sovereignty they aim to preserve.

Embedding agents in data cooperatives could improve collective protections but raises questions about whose risk tolerance dominates when preferences conflict, and whether shared norms reduce individual agency or enhance it through collective bargaining power.


\bibliographystyle{ACM-Reference-Format}
\bibliography{bibliography}
\end{document}